\documentstyle[aps,twocolumn,prl,graphicx]{revtex}  

\begin{document}
\draft
%
\twocolumn[\hsize\textwidth\columnwidth\hsize\csname 
@twocolumnfalse\endcsname 
%

\title{Heavy fermion superconductivity 
and magnetic order in non-centrosymmetric $\rm CePt_3Si$}

\author{E.~Bauer$^a$, G. Hilscher$^a$, H. Michor$^a$, Ch. Paul$^a$,
E.W. Scheidt$^b$, \\   A. Gribanov$^c$, Yu. Seropegin$^c$,  
H. No$\rm \ddot e$l$^d$, M. Sigrist$^e$ and P. Rogl$^f$}

\address{$^a$Institut f\"ur Festk\"orperphysik,
Technische   Universit\"at Wien, A-1040 Wien, Austria}
\address{$^b$Chemische Physik und Materialwissenschaften, 
Universit\"at Augsburg, D - 86159 Augsburg, Germany}
\address{$^c$Dept. of Chemistry, Moscow State University, 
Moscow, Russia}
\address{$^d$Laboratoire de Chimie du Solide et 
Inorganique Mol\'eculaire, Universit\'e de Rennes I,  F-35042 Rennes, France}
\address{$^e$Theoretische Physik, ETH-H\"onggerberg, 8093 Z\"urich,
  Switzerland} 
\address{$^f$Institut f\"ur Physikalische Chemie,
Universit\"at Wien, A-1090 Wien, Austria}

\date{\today}
\maketitle

\begin{abstract}
$\rm CePt_3Si$ is a novel heavy fermion superconductor, crystallising in 
the $\rm CePt_3B$ structure as a tetragonally distorted low 
symmetry variant of the $\rm AuCu_3$ structure type. 
$\rm CePt_3Si$ exhibits antiferromagnetic order at $T_N \approx 2.2$~K
and enters into a heavy fermion superconducting state at $T_c \approx 0.75$~K.
Large values of 
$H_{c2}' \approx -8.5$~T/K and $H_{c2}(0) \approx 5$~T
refer to heavy quasiparticles forming Cooper pairs. 
Hitherto, $\rm CePt_3Si$  is the first 
heavy fermion superconductor  without a center of symmetry.

\end{abstract}


%
] 
\narrowtext
%


Correlation effects among electrons belong to the key causes 
for the occurrence of extraordinary properties of solids at low temperatures.
The most exciting phenomenon in this respect is superconductivity.
Both high temperature - 
and heavy fermion superconductors attracted great interest throughout
the past two decades. 
Most interesting aspects to solve are the specific mechanisms of pairing and
the symmetry of the superconducting condensate \cite{Sigrist}.
While in conventional superconductors the binding
of electrons into Cooper pairs is normally mediated
by phonons, the origin of pairing in some heavy fermion
superconductors is believed to be connected with spin fluctuations, giving 
rise to unconventional superconducting phases (see e.g. Ref.\cite{Mathur}). 

Pronounced electron correlations are generally found in 
systems exhibiting the Kondo effect and thus 
Ce, Yb and U based compounds are certainly candidates for the occurrence of
superconductivity where renormalized quasi-particles
form the Cooper pairs.
While heavy fermion superconductors
are yet missing in Yb based compounds, 
such phases were identified for both Ce and U systems.

Ce-based heavy fermion superconductors, however, are still few in
numbers. Prototypic $\rm CeCu_2Si_2$  exhibits 
superconductivity below $T_c = 0.7$~K \cite{Geibel}.
The application of pressure in the 20 to  30~kbar range 
to members of this structure family such as   $\rm CeCu_2Ge_2$ \cite{Jaccard},
$\rm CePd_2Si_2$ \cite{Grosche} and $\rm CeRh_2Si_2$ \cite{Moshovich}
is sufficient to trigger 
superconductivity as well. 
Very recently, a new class of compounds,
Ce$M$In$_5$, was added  where at ambient conditions heavy 
fermion superconductivity
occurs for $M =$~Co and Ir 
at $T_c  = 2.3$ and 0.4~K, respectively \cite{Petrovich_a,Petrovich_b}. 
Again, pressure initiates  superconductivity, e.g., 
in $\rm CeRhIn_5$ below $T_c^{max} = 2.1$~K \cite{Hegger}. 
The crystal structure of latter compounds
can be considered quasi-two-dimensional
variants of $\rm CeIn_3$  ($\rm AuCu_3$-type).
Cubic $\rm CeIn_3$ becomes
superconducting at   $\sim$25~kbar 
\cite{Mathur}.

The aim of the present paper is to report on the discovery
of both, heavy fermion 
superconductivity and long range magnetic order in the compound
$\rm CePt_3Si$, to evaluate parameters characterising 
the superconducting state and to discuss possible pairing scenarios.

\begin{figure}
\centerline{\includegraphics[width=7cm]{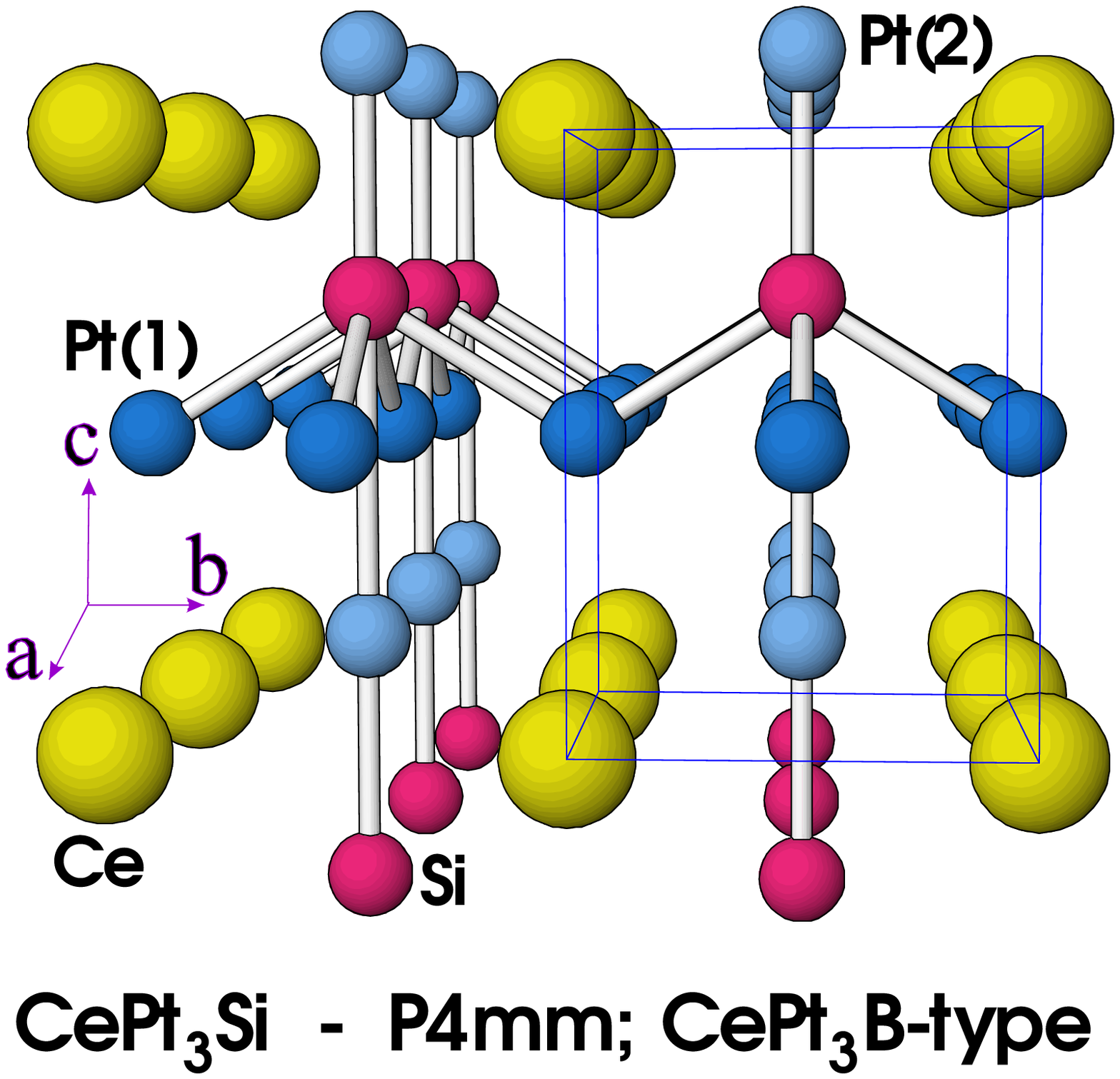}}
\caption{Crystal structure 
of $\rm CePt_3Si$. The bonds indicate the pyramidal coordination
$\rm [Pt_5]Si$ around the Si atom. Origin shifted by (0.5,0.5,0.8532) for 
convenient comparison with parent $\rm AuCu_3$ structure.}
\label{fig0}                          
\end{figure}

$\rm CePt_3Si$ was prepared by argon arc melting and subsequent
heat treatment under high vacuum at $870^{\circ}$~C for three weeks.
Crystal structure 
was determined from Kappa-CCD single crystal X-ray data 
and found to  be tetragonal, 
space group $P4mm$ (No. 99),  
isotypic with the ternary boride $\rm CePt_3B$ \cite{Sologub,Sullow}
(see Fig. \ref{fig0}).
Crystallographic data (standardized) are: 
$a = 0.4072(1)$~nm and $c = 0.5442(1)$~nm; 
Ce in site 1(b)  at $0.5,0.5,0.1468$; Pt(1) in 2(c)  at
$0.5,0,0.6504$, Pt(2) in 1(a) at $0,0,0$ (fixed) and
Si in site 1(a) at $0, 0, 0.4118$.

$\rm CePt_3Si$ derives from 
hypothetical $\rm CePt_3$ with cubic 
$\rm AuCu_3$ structure by filling the void with Si, which 
in addition causes a tetragonal distortion of the unit cell
with $c/a = 1.336$. Filling of voids in structures
with large cages may dramatically influence physical properties
as evidenced in detail for skutterudites $\rm RETM_4X_{12}$,
(RE, rare earths; TM, transition metals;
X, pnicogen) \cite{Bauer}. Of particular importance with respect 
to superconductivity is the lack of a center of inversion
in the crystal structure of $\rm CePt_3Si$.

Electron microprobe analyses and 
Rietveld refinements  revealed phase purity of the polycrystalline material
used for bulk property measurements \cite{probe}.
The characterisation of the paramagnetic state of
$\rm CePt_3Si$ from susceptibility measurements
(not shown here)
reveals a Curie Weiss behaviour
with an effective Ce moment $\mu _{eff} = 2.54~\mu _B$ and a
paramagnetic Curie temperature $\theta _p = -46$~K
(data for the fit are taken from the interval $100 \leq T \leq 300$~K). 
While the former quantity 
indicates a rather stable $3^+$ state of Ce at high temperatures,
the large negative $\theta_p$ value refers 
to pronounced antiferromagnetic interactions. 
In terms of Kondo type interactions,
neglecting crystal field effects,
the Kondo temperature would follow from 
$T_K \approx \vert \theta_p \vert /4 \approx 11$~K \cite{Hewson}.

\begin{figure}
\centerline{\includegraphics[width=7cm]{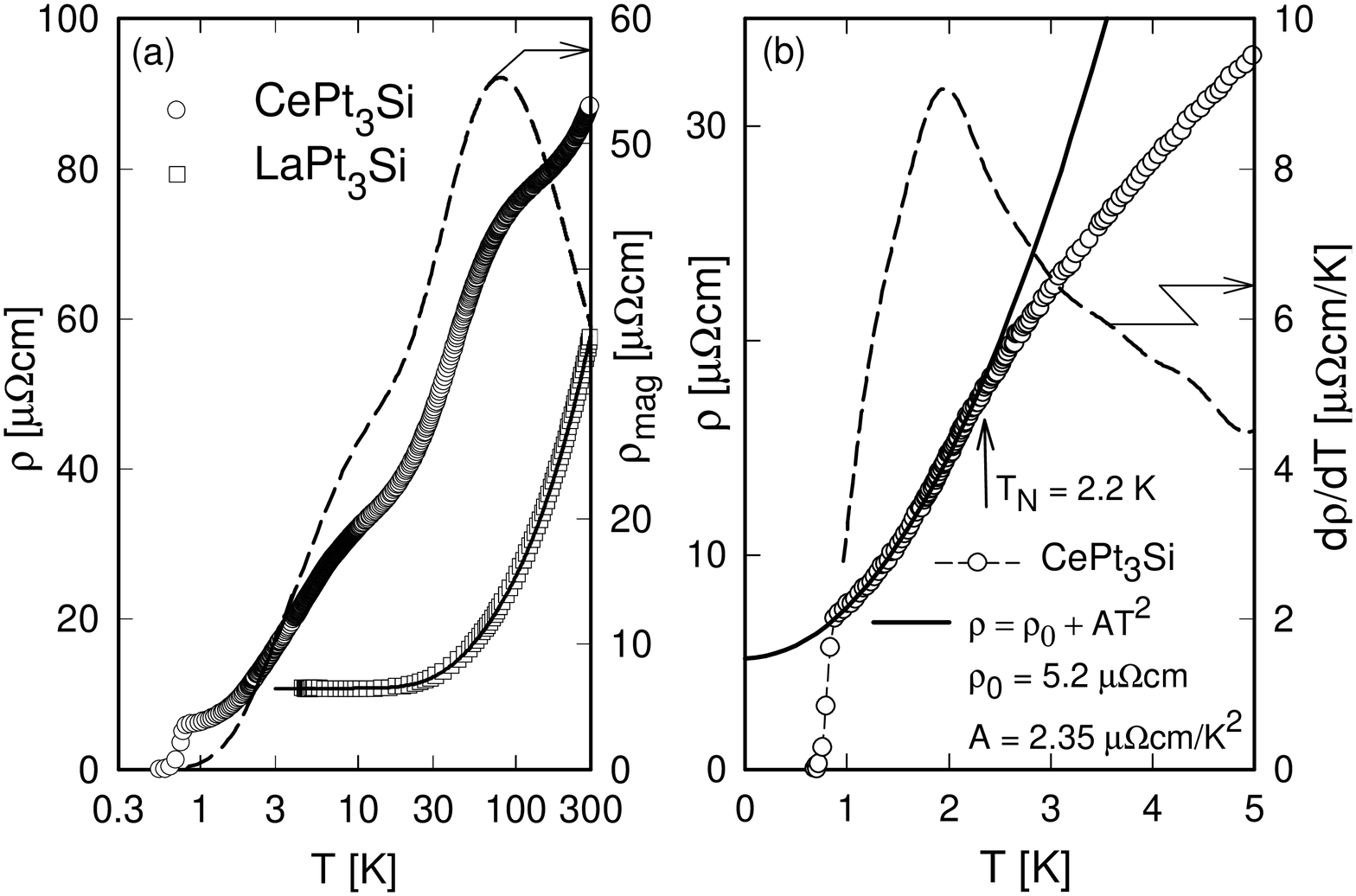}}
\caption{(a): Temperature  dependent electrical resistivity
$\rho$ of $\rm CePt_3Si$ and $\rm LaPt_3Si$ plotted on a 
logarithmic temperature scale. The magnetic contribution 
$\rho_{mag}(T)$ (dashed line) refers to the right axis. 
(b): Low temperature details of $\rho(T)$ of $\rm CePt_3Si$.
The solid line is a least squares fit (see text) and the dashed line 
shows d$\rho (T)/$d$T$.} 
\label{fig2}                        
\end{figure}

Evidence of superconductivity in $\rm CePt_3Si$ is found 
from resistivity
measurements, $\rho (T)$, displayed in Fig. \ref{fig2} together with data of 
isostructural non-magnetic $\rm LaPt_3Si$.
$\rho(T)$ of $\rm CePt_3Si$ drops to zero, 
resulting in $T_c^{mid} = 0.75$~K. 
At high temperatures,  $\rho (T)$ is characterised by a negative 
logarithmic term, followed by pronounced curvatures at about 75~K
and 15~K, which may reflect crystal electric field effects in the 
presence of Kondo type interactions.
Further evaluation of $\rho (T)$ requires knowledge
of the phonon contribution which may be taken from homologous 
and isotypic $\rm LaPt_3Si$ (a = 0.4115(1)~nm, c = 0.5438(2)~nm). 
$\rm LaPt_3Si$ is metallic in the temperature range measured
and is simply accounted for in terms of the Bloch
Gr\"uneisen model with a  
Debye temperature $\theta_D \approx 160$~K [solid line, Fig. \ref{fig2}(a)]. 
The magnetic contribution to the resistivity, $\rho_{mag}(T)$,
is obtained by 
subtracting from the total measured effect of $\rm CePt_3Si$,
both the phonon part 
(taken from $\rm LaPt_3Si$) as well as the respective 
residual resistivities.  $\rho_{mag}(T)$ 
exhibits a  distinct logarithmic contribution
for $T > 100$~K; the maximum around
80~K may indicate the overall crystal field splitting 
of the $j = 5/2$ Ce $4f^1$ state
[dashed line, Fig. \ref{fig2}(a)]. 

Fig. \ref{fig2}(b) illustrates low temperature features of the 
electrical resistivity of $\rm CePt_3Si$. Besides the onset
of superconductivity, there is a distinct change of the slope 
in $\rho (T)$ around 2~K, which becomes more evident from a 
$\rm d \rho / d$$T$ plot [right axis, Fig. \ref{fig2}(b)].  
In the context of specific heat data (see below), 
this anomaly is interpreted as a signature of an initiation
of long range magnetic order.
A least squares fit according to $\rho = \rho_0 + A T^2$, 
reveals the residual resistivity $\rho_0 = 5.2~\mu\Omega$cm
and a material dependent constant $A = 2.35~\mu\Omega$cm/K$^2$.

\begin{figure}
\centerline{\includegraphics[width=7cm]{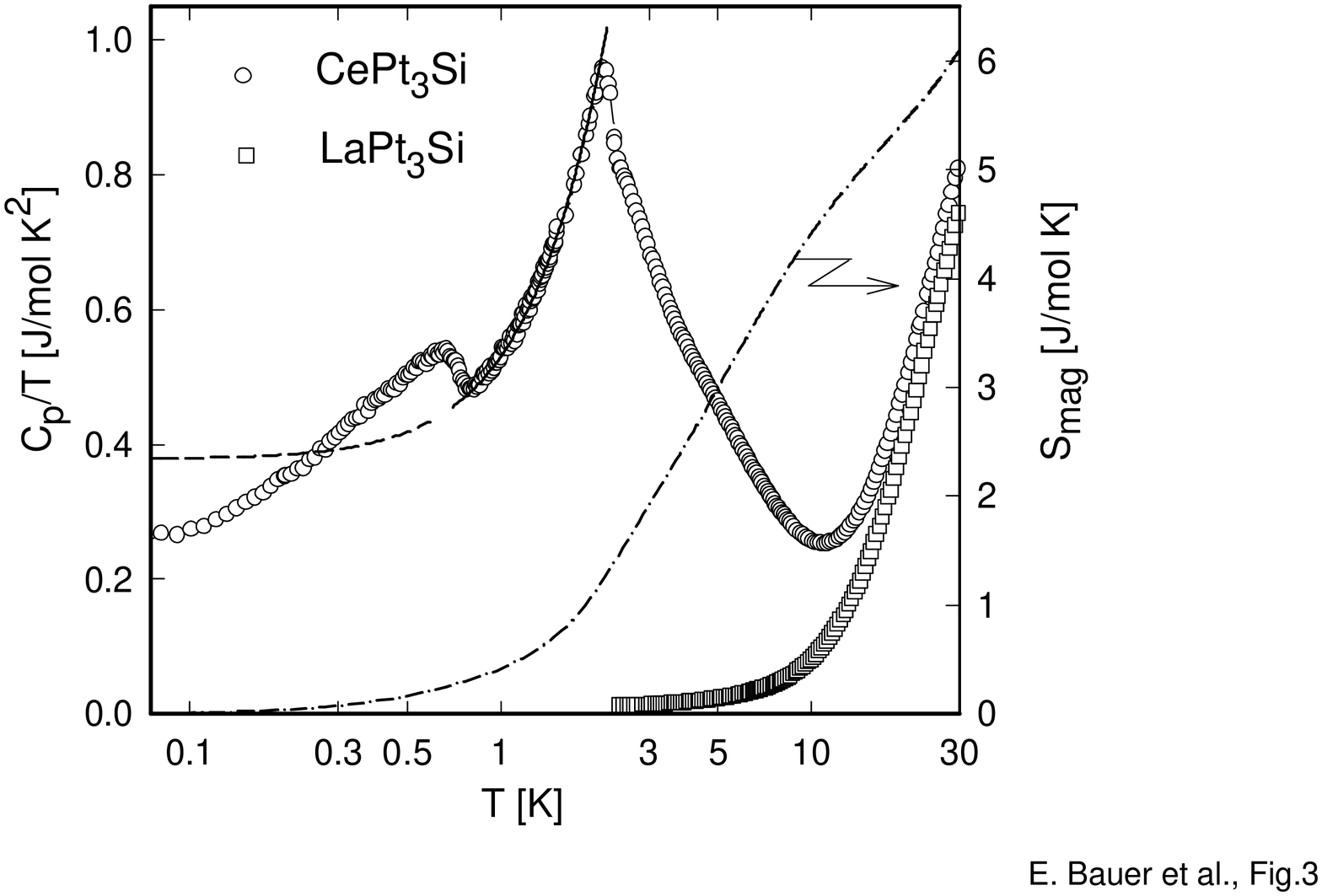}}
\caption{Temperature  dependent specific heat $C_p$
of $\rm CePt_3Si$ and $\rm LaPt_3Si$ plotted as $C_p/T$ vs. $\ln T$.
The solid and the dashed lines are least squares fits (see text). 
Right axis: Temperature dependent magnetic entropy $S_{mag}$
of $\rm CePt_3Si$ (dashed-dotted line).} 
\label{fig3}
\end{figure}

Corroboration of bulk superconductivity 
can be read off from specific heat measurements, $C_p (T)$.
Results are shown in Fig. \ref{fig3} as $C_p/T$ vs. $\ln T$. 
Data of $\rm LaPt_3Si$ are added for comparison.
Standard analysis for the latter yields $\gamma = 9~$mJ/molK$^2$,
and $\theta _ D^{LT} = 255$~K. 
Two different features are obvious for $\rm CePt_3Si$
(from higher to lower temperatures):
i) a distinct $\lambda$-like anomaly at about 2~K
and ii) a jump of the specific heat around 0.7~K.

The $\lambda$-like anomaly of the specific heat 
at 2.2~K signifies the onset of long range
magnetic order. A $T^3$-dependence
of $C_p(T)$ well below the transition may 
characterise antiferromagnetic ordering (solid line, Fig. \ref{fig3}).
Antiferromagnetic order 
also follows from a field driven shift of the $\lambda$-like anomaly
towards lower temperatures and is confirmed by
the absence of spontaneous magnetisation.
Above  $T_N$, $C_p/T$ of $\rm CePt_3Si$ exhibits 
an extended -- almost logarithmic -- tail, 
indicative of pronounced spin fluctuations in the paramagnetic state.
The anisotropy of the crystal structure  ($c/a \simeq 1.34$)
is supposed to be an additional cause
for such enhanced short range magnetic correlations.

The magnetic entropy $S(T)$ was derived from a comparison of $C_p(T)$
of both $\rm CePt_3Si$ and $\rm LaPt_3Si$. Results are shown as 
dashed-dotted line in Fig. \ref{fig3}.
The entropy gain of about $\rm 0.22 \times R \ln 2$ at $T = T_N$ 
is well below that associated with the lifting of the degeneracy
of the ground state doublet and 
suggests ordering with substantially reduced magnetic 
moments. Involving Kondo type interactions 
as possible mechanism for the above observation would
reflect a characteristic temperature $T_K$ of about 10 to 15~K. 
$R\ln 2$ is reached around 25~K only.

The most interesting feature, the anomaly around 0.7 K, indicates the
transition  into a superconducting phase,
in agreement with the above $\rho(T)$ data. 
In order to accurately determine $T_c$, the standard
procedure with an idealised jump at $T_c$ is applied,
yielding $T_c = 0.75$~K.    
Some estimation of the normal 
state Sommerfeld value $\gamma_n$ may be obtained from an
extrapolation of the $T^3$-dependence 
(dashed line, Fig. \ref{fig3}), 
arriving at $\gamma_n \approx 0.39$~J/molK$^2$. 
This extrapolation  primely satisfies the 
basic requirement of entropy balance between the
superconducting  and the normal state. 
The jump of the specific heat $\Delta C_p/T(T = T_c) \approx 0.1$J/molK$^2$, 
allows calculation of the parameter 
$\Delta C_p/(\gamma_n T_c) \approx 0.25$, which is well below 
the figure expected from  BCS theory 
($\Delta C_p/(\gamma T_c) \approx 1.43$). Assessing 
the electronic specific heat coefficient in the superconducting state, 
$\gamma_s \approx 0.18(1)$~J/molK$^2$,   
from the difference between data derived from the $T^3$ 
extrapolation and those estimated at 
low temperature and zero field yields  
$\Delta C_p/(\gamma_s T_c) \approx 0.55$, which is still
under the BCS value. It should be noted that, e.g.,  
$\Delta C_p/(\gamma T_c)$ of the
spin triplet superconductor $\rm Sr_2RuO_4$
is similarily downsized \cite{Makenzie}.

\begin{figure}
\centerline{\includegraphics[width=7cm]{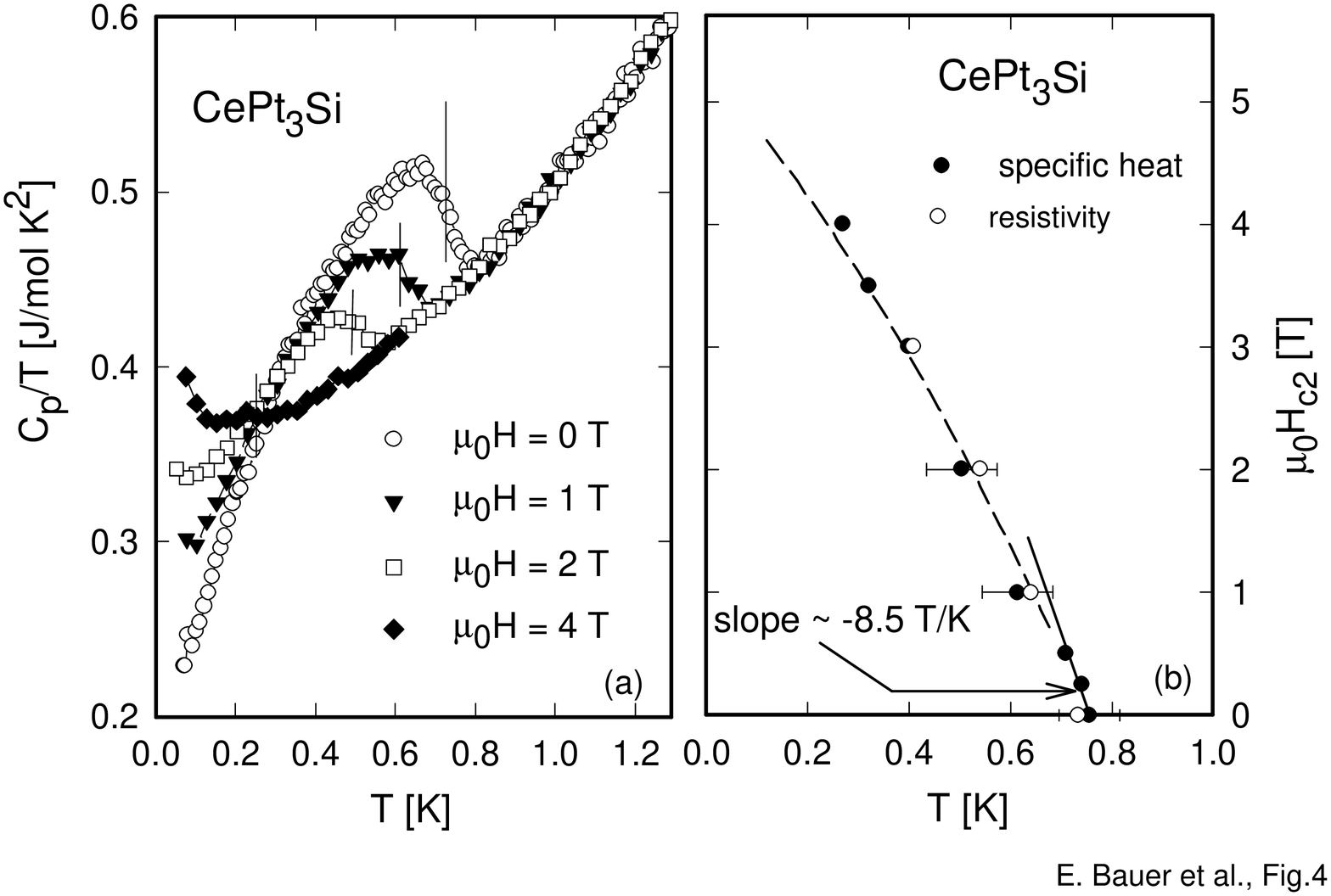}}
\caption{(a): Temperature  dependent specific heat $C_p/T$
of $\rm CePt_3Si$ for various values 
of applied fields; vertical lines indicate the SC transition. 
(b): Temperature dependence of
the upper critical field $H_{c2}$ deduced from specific heat and 
resistivity measurements. The solid straight line yields 
$H_{c2}' \approx - 8.5$T/K and the dashed line is a guide to the
eyes.  } 
\label{fig4}
\end{figure}

The specific heat of $\rm CePt_3Si$ for various external fields
is plotted in Fig. \ref{fig4}(a). The application of magnetic fields
reduces $T_c$, giving rise to a rather large change of 
$d H_{c2} /dT \equiv H_{c2}' \approx - 8.5$~T/K, 
in excellent agreement with the respective
data from electrical resistivity [see Fig. \ref{fig4}(b)].
Extrapolation of the field dependent transition 
temperatures towards zero  yields $H_{c2}(0) \approx 5$~T.
Furthermore, an estimation of the Sommerfeld coefficient
for high fields results in 0.36~J/molK$^2$,
in fair agreement with the value gained from 
an extrapolation of normal state, zero field data (see Fig. \ref{fig3}).
The upturn of $C_p/T$ at lowest temperatures,
reenforcing in magnetic fields,
is primarily associated with the nuclear contribution of $\rm ^{195}Pt$.

By analogy to commonly employed practice concerning heavy fermion 
superconductors, we estimate in the following a number of parameters 
from an analysis of superconducting and normal state properties
in terms of the BCS theory
\cite{Tinkham,Rauchschwalbe} assuming a spherical Fermi surface and
incorporating clean and dirty limit terms.
Starting parameters are $\gamma_s = 0.18(2)$~J/molK$^2$
(as a lower limit), $H_{c2}' = - 8.5$~T/K
and $\rho_0 = 5.2~\mu\Omega$cm. Although deviations from a spherical 
Fermi surface are expected  for tetragonal $\rm CePt_3Si$, 
reasonable physical parameters can be anticipated
(compare, e.g. Ref. \cite{Rauchschwalbe,Movshovich}).

The effective Fermi surface $S_s$ is computed 
(Eqn. 2 in Ref. \cite{Rauchschwalbe}) as  
$S_s^{cl} \simeq 3.7 \times 10^{20}$~m$^{-2}$ within the clean - and 
$S_s^{dl} \simeq 3.5 \times 10^{20}$~m$^{-2}$ for the dirty limit.
Considering the dirty limit only, $H_{c2}'(calc) = - 0.77$~T/K,
a value rather low with respect to the experimentally derived
slope $H_{c2}' = - 8.5$~T/K. This indicates that $\rm CePt_3Si$ is not
a typical dirty limit superconductor. 
Irrespective of all the shortcomings present, further calculations 
are carried out using clean limit parameters.

Combining the Fermi surface with $\gamma_s$
gives the Fermi velocity $v_F \simeq 5300$~m/s
and in the context of the residual resistivity, $\rho_0 = 5.2~\mu \Omega$cm,
a mean free path $l_{tr} \simeq 8 \times 10^{-8}$~m 
can be derived.
The coherence length $\xi _0$ for $T \rightarrow 0$  was obtained from 
two independent relations.
One follows from the BCS equation,
 $\xi _0 = 0.18 \hbar v_F / (k_B T_c) \simeq 9.7 \times 10^{-9}$~m.
 A second expression stems from the well known formula 
$\mu_0 H_{c2} = \Phi _0 / (2 \pi \xi _0 ^2 )$, yielding 
$\xi _0  \simeq 8.1 \times 10^{-9}$~m, in reasonable agreement with the former. 
Evaluation of the Ginzburg Landau parameter $\kappa _{GL} = \lambda / \xi$
requires the knowledge of the thermodynamic critical field 
$\mu_0H_c(0)\simeq 26$\,(2)\,mT, which 
is calculated from the free energy difference 
between the superconducting  and the normal state:
$\Delta F (T) = F_n - F_s = \mu_0 H_c^2 (T)/2
 = \int_{T_c}^T \int_{T_c}^{T^{\prime }} 
 \frac{(C_s - C_n) }{T^{\prime \prime }}\,dT^{\prime \prime }dT^{\prime }$.
$C_s$ is obtained from 
the zero field specific heat and $C_n$ is taken from the 
$T^3$ extrapolation as indicated by the dashed line in Fig. \ref{fig3}.
With $H_{c2}(0) \approx 5$~T one derives a value for 
$\kappa_{GL} = H_{c2}(0) /(\sqrt 2 H_c) \simeq 140$ which,
in turn, determines the London penetration depth
$ \lambda _L (T \rightarrow 0) \simeq 1.1 \times 10^{-6}$~m.

Evaluating Eqn. A.13 of Ref. \cite{Rauchschwalbe} with 
$\rho_{max} \approx 100~\mu \Omega$cm yields 
$S_{hT} \simeq 3.1 \times 10^{21}$~m$^{-2}$, the 
Fermi surface at elevated temperatures. 
The discrepancy between $S_{s}$ and  $S_{hT}$ suggests
that only a minor part of the Fermi surface is involved in forming 
Cooper pairs while the major part engages in normal state 
magnetic correlations. This finding seems to be convincingly 
supported from the lessened value of  
$\Delta C_p/(\gamma T_c)$.  
In terms of the coexistence of both superconductivity
($T_c = 0.75$~K) and long range magnetic order ($T_N = 2.2$~K), 
the downsized specific heat jump at 
$T_c$ may explain, at least partly, that 
the Fermi surface is likely to be subdivided into a superconducting
part (related to $\gamma_s$) and a normal state region.

To classify the behavior of $\rm CePt_3Si$ within a wider framework,
we adopt a generic phase diagram which has been observed for various
Ce-based systems as well as for $3d$ systems such as $\rm MnSi$ 
\cite{Pfleiderer}. 
With an external control parameter $\delta$, like doping or
pressure, the system may be shifted towards $T_{mag} = 0$,
defining the quantum critical point (QCP) of magnetic order. 
In several cases a dome of superconductivity has been discovered in a
region around a QCP, e.g. in $\rm CeIn_3$ \cite{Mathur} or
$\rm CePd_2Si_2$ \cite{Grosche}. The superconducting instability
occurs on the background of fluctuation-induced non-Fermi liquid
behavior. From this point of view we would place $\rm CePt_3Si$ slightly away
from the QCP towards the magnetically ordered region.
On cooling, the system  undergoes 
successive phase transitions to magnetic order and superconductivity.

The discussion of pairing symmetry in $\rm CePt_3Si$ raises an
interesting problem. Here superconductivity occurs within a
magnetically ordered 
phase. This coexistence alone does not imply unconventional pairing. 
The strong electron correlation effects, however, give very likely 
rise to pairing with a higher angular momentum.  
An important aspect of $\rm CePt_3Si$ is the lack of an inversion
center in the crystal structure. It is believed that this excludes
spin-triplet pairing, since in the absence of inversion symmetry
the necessary set of degenerate electron states cannot be provided for this
type of pairing \cite{Anderson}. This argument was used to explain the
absence of superconductivity close to the quantum critical point in
weakly ferromagnetic MnSi \cite{Saxena}, although this material exists as ultra
pure single crystal excluding pair breaking
scattering \cite{Pfleiderer}. 
The upper critical field $ H_{c2}(0) $ in $\rm CePt_3Si$ is
surprisingly high, and exceeds the Pauli-Clogston limiting 
field, which in a simple-minded approach is given by
$ H_p \approx \sqrt{2} \Delta / \mu_B \sim 3 k_B T_c / \mu_B
\approx 2 T < H_{c2}(0) $, implying 
per definition  an effective electron g-factor of 2.
If this rough estimate holds for $\rm CePt_3Si$, 
the high $ H_{c2}(0) $ would then be incompatible with spin-singlet
pairing and rather signals spin-triplet superconductivity.
The apparent conflict may be resolved by noticing that the lack of
inversion symmetry is not excluding spin triplet pairing completely. 
In our case the space group P4mm involves the absence of the mirror plane 
$ z \to -z $ which yields a Rashba-type of spin orbit coupling
and leaves the triplet (equal-spin) pairing 
state $ {\bf d}({\bf k}) = \hat{\bf x} k_y - \hat{\bf y} k_x $ 
(irreducible representation $ A_{2u}
$ of  $D_{4h}$) as a possible pairing state
\cite{Agterberg,Gorkov}.

In summary,  $\rm CePt_3Si$ is a heavy fermion compound
undergoing both a magnetic transition at $T_N = 2.2$~K
and a superconducting transition at $T_c = 0.75$~K. 
The ratio $\l _{tr} / \xi _0 \simeq 8$ 
is indicative of a clean limit superconductor.
Thermodynamic data derived for $\rm CePt_3Si$ suggest
that the Cooper pairs are formed of heavy quasiparticles. 
The pairing is likely affected by the absence of an inversion center. 
Relatively large $H_{c2} $ might be a hint for the presence of 
spin-triplet pairing. While still speculative, the prospect of
spin triplet pairing in a system without inversion symmetry is
an exciting issue for future studies. 
 
We are grateful to D.F. Agterberg for
helpful discussions.
This work was supported by the Austrian FWF  P16370, 15066,
by INTAS project 234, by the Deutsche Forschungsgemeinschaft (DFG)
SFB 484 (Augsburg), by the ESF project FERLIN and the Swiss
Nationalfonds.



\begin{thebibliography}{999}


\bibitem{Sigrist} M. Sigrist and K. Ueda, Rev. Mod. Phys. {\bf 63}, 239 (1991).

\bibitem{Mathur} N.D. Mathur et al., 
Nature {\bf 394}, 39 (1998). 

\bibitem{Geibel} F. Steglich et al., Phys. Rev. Lett {\bf 43}, 1892 (1979).

\bibitem{Jaccard} D. Jaccard et al., Phys. Lett. A {\bf 163}, 475 (1992).

\bibitem{Grosche} F.M. Grosche et at., Physica B {\bf 224}, 50 (1996).

\bibitem{Moshovich} R. Movshovich et al., Phys. Rev. B {\bf 53}, 8241 (1996).  

\bibitem{Petrovich_a} C. Petrovic et al., Europhys. Lett. {\bf 53}, 354 (2001).

\bibitem{Petrovich_b} C. Petrovic et al., 
J. Phys.: Cond. Mat. {\bf 13}, L337 (2001).

\bibitem{Hegger} H. Hegger et al., Phys. Rev. Lett. {\bf 84}, 4986 (2000).


\bibitem{Sologub} O. Sologub, et al.,
J. Alloys Compounds {\bf 337} (2002) 10.

\bibitem{Sullow}
$\rm CePt_3B$ was first observed to form a (centro-symmetric!)
tetragonal structure 
[Physica B {\bf 199 -200}, 644 (1994)].

\bibitem{Bauer} E. Bauer et al., Acta Phys. Pol. B {\bf 34}, 595 (2003).


\bibitem{probe} 
Same sample was used for all measurements, 
and several samples investigated proved full reproducibility.


\bibitem{Hewson} A. Hewson, {\it The Kondo Problem to Heavy Fermions},
Cambridge University Press, 1993. 


\bibitem{Makenzie} A.P. Makenzie and Y. Maeno, 
Rev. Mod. Phys. {\bf 75}, 657 (2003).



\bibitem{Tinkham} See, for example 
M. Tinkham, {\it Introduction to Supercondcutivity},
McGraw-Hill, New York, 1975


\bibitem{Rauchschwalbe} U. Rauchschwalbe, Physica {\bf 147 B}, 1 (1987).

\bibitem{Movshovich} R. Movshovich, et al., Phys. Rev. Lett. 
{\bf 86}, 5152 (2001).


\bibitem{Pfleiderer} Ch. Pfleiderer et al., 
Phys. Rev. B {\bf 55}, 8330 (1997).


\bibitem{Anderson} P.W. Anderson, Phys. Rev. B{\bf 30}, 4000 (1984). 


\bibitem{Saxena} S.S. Saxena et al.,
Nature, {\bf 406}, 587 (2000). 


\bibitem{Agterberg} D.F. Agterberg and M. Sigrist, in preparation. 

\bibitem{Gorkov} L.P. Gorkov and E.I. Rashba, Phys. Rev. Lett. {\bf
  87}, 037004 (2001). 

\end{thebibliography}
\end{document}